\pdfoutput=1
\documentclass[11pt]{article}

\usepackage[final]{acl}

\usepackage{times}
\usepackage{latexsym}
\usepackage{amsmath}
\usepackage{graphicx}
\usepackage{booktabs}
\usepackage{tabularx}

\usepackage[T1]{fontenc}
\usepackage[utf8]{inputenc}
\usepackage{multirow}
\usepackage[table,xcdraw]{xcolor}

\usepackage{microtype}
\usepackage{colortbl}
\usepackage{inconsolata}

\newcommand{\ours}{\textsc{ConvInv} }
\newcommand{\gc}{\cellcolor[HTML]{E2FFE2 }}
\newcommand{\rc}{\cellcolor[HTML]{FFCCC9 }}
\title{Interpreting Conversational Dense Retrieval by Rewriting-Enhanced Inversion of Session Embedding}

\author{Yiruo Cheng, Kelong Mao, Zhicheng Dou\thanks{Corresponding author.} \\
       Gaoling School of Artificial Intelligence, Renmin University of China \\
       \texttt{\{chengyr,mkl,dou\}@ruc.edu.cn}  }

\begin{document}
\maketitle
\begin{abstract}
Conversational dense retrieval has shown to be effective in conversational search. However, a major limitation of conversational dense retrieval is their lack of interpretability, hindering intuitive understanding of model behaviors for targeted improvements. This paper presents \ours, a simple yet effective approach to shed light on interpretable conversational dense retrieval models. 
\ours transforms opaque conversational session embeddings into explicitly interpretable text while faithfully maintaining their original retrieval performance as much as possible.
Such transformation is achieved by training a recently proposed Vec2Text model~\cite {vec2text} based on the ad-hoc query encoder, leveraging the fact that the session and query embeddings share the same space in existing conversational dense retrieval.
To further enhance interpretability, we propose to incorporate external interpretable query rewrites into the transformation process. Extensive evaluations on three conversational search benchmarks demonstrate that \ours can yield more interpretable text and faithfully preserve original retrieval performance than baselines. Our work connects opaque session embeddings with transparent query rewriting, paving the way toward trustworthy conversational search.
Our code is available at \href{https://github.com/Ariya12138/ConvInv}{this repository}.
\end{abstract}

\section{Introduction}

\begin{figure}[ht]
  \centering
  \includegraphics[width=0.5\textwidth]{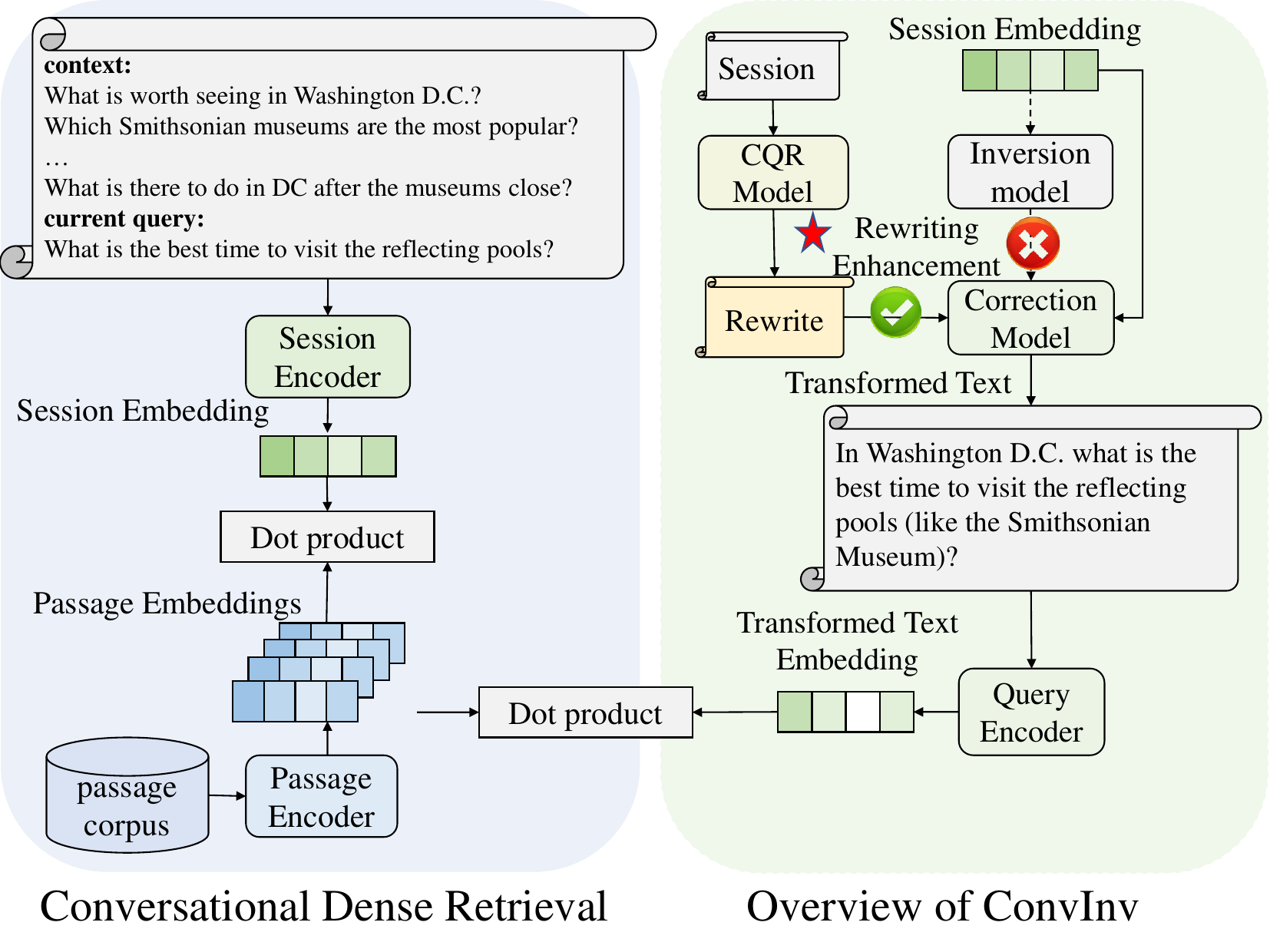}
  \caption{The blue section on the left signifies the conversational dense retrieval, and the green section on the right provides an overview of \textsc{ConvInv}.}
  \label{overview}
\end{figure}

With the rapid development of language modeling, conversational search has emerged as a novel search paradigm and is garnering more and more attention. Different from the traditional ad-hoc search paradigm characterized by keyword-based queries and “ten-blue” links~\citep{Yu:CQR}, conversational search empowers users to interact with the search engine through multi-turn natural language conversations to seek information, which brings a more intuitive and efficient search experience~\citep{Mao:ConvTrans, microsoft22_neural_CIR_survey,LLM4IRSurvey}.

% The support for multi-turn natural language conversations allows the search engine to better deal with users' complex information needs.

In conversational search, the system input is a multi-turn natural language conversation, which may have many linguistic problems such as omissions, co-references, and ambiguities~\citep{Radlinski}, posing great challenges for accurately grasping the user's real information needs.
Recently, conversational dense retrieval (CDR)~\citep{Yu:CDR,emnlp21_CQE,emnlp22_saving_shortcut_cdr,Mao:COTED, emnlp22_CRDR, Mo:CQRKDD,CDMLZ2024ACL}, which directly encodes the whole conversational search session and the passages into a unified embedding space to perform matching, has shown to be a promising method to solve this complex search task.
Compared to another type of method: conversational query rewriting (CQR)~\citep{Sheng-Chieh:CQR, Vakulenko:CQR, conqrr, ConvGQR}, which is a two-step method that first reformulates the search session into a decontextualized query rewrite and subsequently inputs this rewrite into existing ad-hoc search models for search, the end-to-end CDR models can be directly optimized towards better search effectiveness~\citep{Yu:CDR} and is more efficient as it avoids the extra latency caused by the rewriting step.

% While there existing omission, references to previous turns and ambiguities~\citep{Radlinski} in natural language conversations, which poses challenges for conversational search to accurately understand users’ information need. There are mainly two methods to address this challenge: conversational query rewrite(CQR) and conversational dense retrieval(CDR). Conversational Dense Retriever encodes the entire context into embeddings that convey users’ information need, and then calculates similarity with the embeddings of the passages~\citep{Yu:CDR}. In comparison to another intuitive approach, the CQR method, the CDR model excels in direct optimization based on downstream retrieval signals, manifesting promising retrieval effectiveness. 

However, a notable drawback of conversational dense retrieval is that it inherently lacks interpretability~\citep{LeCoRE}.
By encoding conversations into dense vector embeddings rather than readable text, it becomes opaque how these CDR models comprehend search intent.
The absence of interpretability becomes a severe obstacle for developers to comprehend the reasons behind the search results, hindering effective and targeted enhancements to the bad cases of the models~\citep{LeCoRE,acl23_findings_edircs}.
Moreover, the absence of interpretability poses challenges in identifying and addressing potential biases or errors within the models, which could lead to unfair or misleading search results without the possibility of timely correction.

In this paper, we present \ours: a simple and effective approach aiming to shed light on the opacity problem of conversational dense retrieval.
\ours demystifies the opaque conversational session embeddings by transforming them into explicitly interpretable text while faithfully maintaining their retrieval performance as much as possible.
This transformation allows us to intuitively decipher the characteristics of behaviors of different conversational dense retrieval models. 

Figure~\ref{overview} provides an overview of \textsc{ConvInv}. Specifically, our approach is based on the recently proposed Vec2Text~\citep{vec2text}, which is a powerful method that can invert any text embedding into its original text given the corresponding text encoder.
However, inverting the session embedding into the original session is meaningless as it brings no interpretability.
We adapt Vec2Text to suit our interpretable inversion of conversational session embedding by taking specific advantage of how the conversational session encoders are trained: the session encoder starts from an ad-hoc query encoder and the passage encoder is frozen during the training. This makes the session and query embeddings finally share the same embedding space for retrieval.
Therefore, we propose to train a Vec2Text model based on the ad-hoc query encoder to transform the session embedding so that the transformed text is different from the original session, but also maintains a similar retrieval performance when encoding it with the ad-hoc query encoder.
To further enhance the interpretability of the transformed text, we directly incorporate well-interpretable external query rewrites into the Vec2Text transformation process, effectively guiding it to yield more interpretable text. 

We conduct extensive evaluations on three conversational search benchmarks.
Compared to baselines, the proposed \ours can transform conversational session embeddings into more interpretable text as well as faithfully restore the original retrieval performance of the session embeddings.

In summary, the contributions of our work are:

(1) We introduce a simple and effective approach \textsc{ConvInv} to shed light on the interpretability of conversational dense retrieval models by transforming opaque conversational session embeddings into interpretable text as well as
faithfully maintain their original retrieval performance.

(2) We propose to incorporate the query rewrites into the transformation process to effectively enhance the interpretability of the transformed text.

(3) Our work connects opaque session embeddings with transparent query rewriting, paving the way toward trustworthy conversational search.

\section{Related Work}
\subsection{Conversational Search}
% In conversational search, understanding users' query intent and meeting their information needs pose a significant challenge. 
% To address this challenge,
Currently, conversational search primarily relies on two main methods: conversational query rewriting (CQR) and conversational dense retrieval (CDR). CQR~\citep{Yu:CQR,conqrr,Kumar:CQR,Voskarides:CQR,Sheng-Chieh:CQR,Mao:cqrACL,Liu:CQR,Vakulenko:CQR,Vakulenko:CQRECIR,LLM_Know_You,ConvGQR} transforms the whole session into a context-independent query. The generated query rewrites can directly perform ad-hoc retrieval. In contrast, CDR~\citep{Yu:CDR,arxiv24_haconvdr,Krasakis:CDR,Mao:COTED,Mo:CQRKDD,LeCoRE,Mao:ConvTrans,Dai:CDR, ecir_cosplade,arxiv24_chatretriever} aims to train a session encoder that is capable of encoding the conversational context into a high-dimensional space for conducting dense retrieval. However, the session embedding encoded by the conversational query encoder lacks interpretability, hindering developers from comprehending the retrieval results.

\subsection{Interpretable information retrieval}
% With the rapid advancement of dense retrieval methods in the field of information retrieval, interpretability has become an increasingly prominent focus. 
The interpretability issues have increasingly garnered attention within the domain of information retrieval.
%\citep{Adolphs} conducted a study wherein they trained a single-step query decoder. The objective of this decoder is to predict text content from the embeddings of queries. Subsequently, they leveraged this decoder to generate additional data, which was employed in training a new retrieval model. 
\citep{Ram} proposed to interpret the session embeddings from dual encoders by mapping them into the lexical space of the model. ~\citep{LeCoRE} proposed to augment the SPLADE model by incorporating multi-level denoising approaches, which can produce denoised and interpretable lexical session representations. 

%\noindent \textbf{Dense embedding inversion.} 
To explore the intricate interplay between embedded representations and their textual counterparts, a substantial body of research has focused on the task of inverting embeddings to coherent text. Representing the embedding of sentences as the initial token, ~\citep{Haoran} trained a powerful decoder model to decode the entire sequence. ~\citep{vec2text} endeavored to produce text whose embedding closely approximates the given embedding. They achieved this by using the difference between hypothesis embeddings and actual embeddings.
% facilitating discrete updates to the text hypothesis.

\begin{figure*}[ht]
  \centering
  \includegraphics[width=\textwidth]{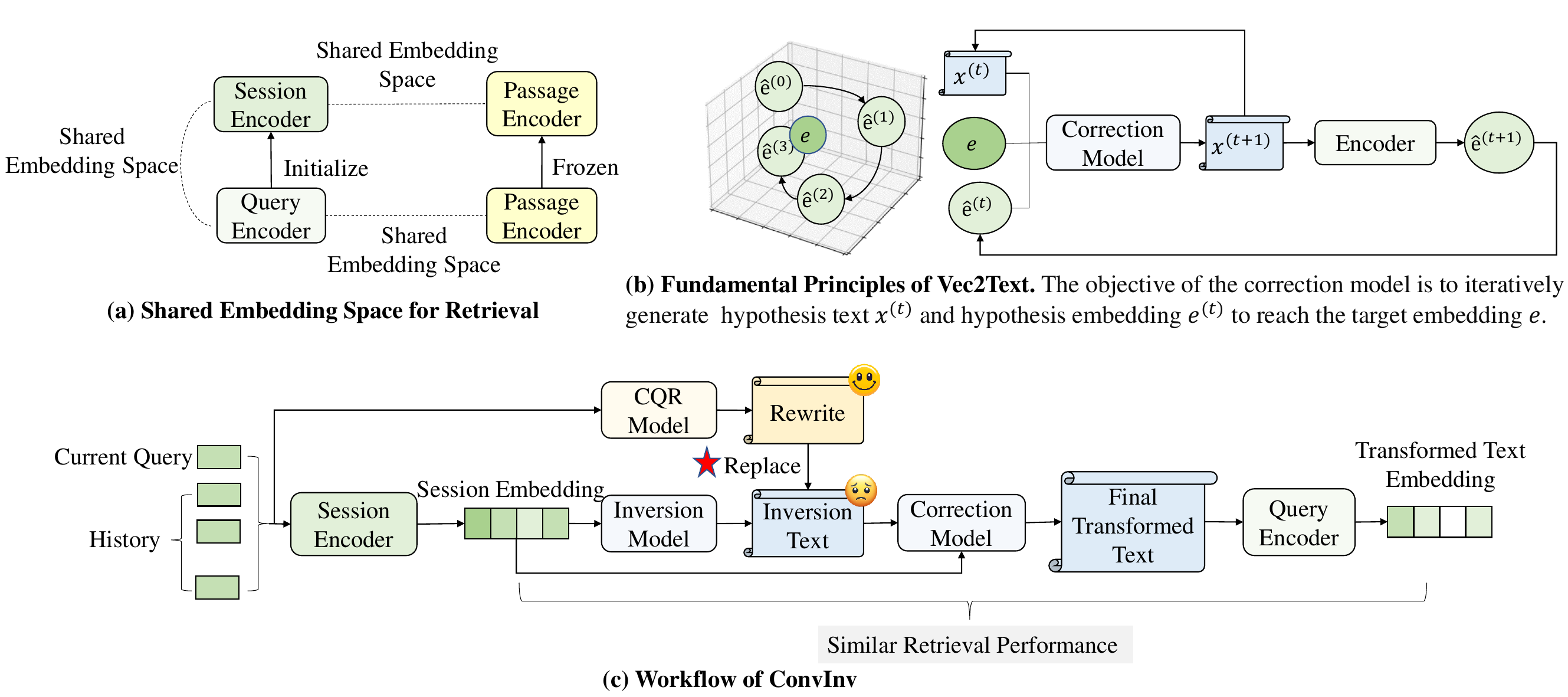}
  \caption{Architecture of our proposed \textsc{ConvInv}.}
  \label{model1}
\end{figure*}

\section{Methodology}

In this work, we present \textsc{ConvInv}, a new approach designed to demystify conversational session embeddings. Our approach focuses on transforming these opaque conversational session embeddings into explicitly interpretable text while maintaining their retrieval performance as much as possible.
\ours aims to bridge the gap between the mysterious nature of dense embeddings and the necessity for clear, understandable insights in conversational search intent analysis.

\subsection{Preliminaries}
% \subsubsection{Conversational Problem Definition}\label{CDR}
% Conversational search is an information retrieval process that revolves around a continuous and interactive conversation between a user and a search agent. 
%\subsubsection{Conversational dense retrieval}
\subsubsection{Conversational dense retrieval} 
Formally, conversational search involves a series of turns $\left \{ \left ( q_{i},a_{i}   \right )  \right \} _{i=1}^{n}  $, where the users express their information needs at $i$-th turn through $q_i$, and the system returns a relevant response $a_i$. 
This paper focuses on the conversational retrieval task, where the goal of conversational search models is to retrieve relevant passages $p$ for the current query $q_i$, considering its historical context $H_i=\left \{ \left ( q_{j}, a_{j}   \right )  \right \} _{j=1}^{i-1}  $. 
The idea of conversational dense retrieval is to jointly map the current query $q_i$ along with the historical context $H_i$ and passages into a unified embedding space, and use the similarity between the session embedding and the passage embedding as the retrieval score:
\begin{eqnarray}
\mathbf{s_i} &=& E_\text{s}(q_i, H_i), \quad \mathbf{p} = E_\text{p}(p), \\
r &=& \text{cos}(\mathbf{s_i}, \mathbf{p}),
\end{eqnarray}
where $E_{\text{s}}$ and $E_{\text{p}}$ are the session and passage encoders, respectively. $\text{cos}$ is the cosine similarity used to compute the retrieval score $r$.

\subsubsection{Task formulation}\label{sec:task_forumation}
The encoded conversational session embedding $\mathbf{s_i}$, while effective, is inherently mysterious and lacks interpretability. Our goal is to transform the session embedding $\mathbf{s_i}$ into an explicit, interpretable text $\hat{q_i}$ while faithfully maintaining the original retrieval effectiveness of the session embedding in $\hat{q_i}$.

\subsection{Our Approach}
To achieve this transformation from session embeddings to interpretable text, we propose a simple yet effective approach, called \ours, which is built upon the Vec2Text model~\citep{vec2text} with tailored adjustments for the interpretation of conversational dense retrieval.
Specifically, our approach has two important steps: (1) Training a Vec2Text model based on the ad-hoc query encoder. (2) Enhancing interpretation with rewriting. 
Figure~\ref{model1} shows an illustration of our approach.

\subsubsection{Training Vec2Text based on Ad-hoc Query Encoder}
Vec2Text~\citep{vec2text} is a recently proposed method for transforming embeddings into text.
Given any text encoder $E$ and a large collection of texts $T = \{t_i\}$ where $t_i$ is a text, a Vec2Text model $\phi$ is trained based on a large number of (embedding, text) pairs (i.e., $\left< E(t_i), t_i\right>$) to learn to invert any text embedding $E(t_i)$ into a text $t^{'}_i$, where $E(t^{'}_i)$ is very similar to $E(t_i)$. 
As reported in their original paper, $\text{cos}(E(t^{'}_i), E(t_i))$ can reach up to 0.99. 
Motivated by the remarkable effectiveness of Vec2Text, we adapt it to suit our interpretable inversion of conversational session embedding by leveraging a specific training characteristic of conversational session encoders: \textit{Shared Embedding Space for Retrieval}. 

\noindent \textbf{Shared embedding space for retrieval.} 
For the training of conversational dense retrievers, it is common to initialize the conversational session encoder and the passage encoder from a pre-trained ad-hoc retriever, and only fine-tune the session encoder while freezing the passage encoder for facilitating the training~\citep{Yu:CDR, emnlp21_CQE, Mao:COTED, Mo:CQRKDD}.
Therefore, we may assume that the session encoder and the ad-hoc query encoder share the same embedding space for retrieval as they share the same passage encoder. This characteristic is ideal for us to achieve more interpretable session embedding inversion as well as maintain its original retrieval effectiveness. 

% We skillfully leverage this training characteristic to achieve interpretable session embedding inversion as well as maintain the original retrieval effectiveness.

% Specifically, there are mainly two paradigms to train conversational session encoders currently. The first is proposed by \citet{Yu:CDR} which employs an ad hoc query encoder $E_q$ as the teacher and learns the student session encoder $E_s$ by mimicking the teacher embeddings on oracle reformulated queries $q^{*}$.
% % using the MSE loss function.
% % \begin{eqnarray}
% % \text{MSE} (E_q(q^{*}_i), E_s(q_i, H_i)).
% % \end{eqnarray}
% The second is to use the classical ranking loss function to maximize the distance between the session and its positive passages and minimize the distance between the session and negative passages.
% % \begin{eqnarray}
% % \text{MSE} (E_q(q^{*}_i), E_s(q_i, H_i)).
% % \end{eqnarray}
% It is important to note that both two paradigms commonly start by initializing the conversational session encoder and the passage encoder using a pre-trained ad-hoc retriever. During the training process, only the session encoder is fine-tuned, while the passage encoder remains unchanged, to facilitate the training~\cite{Yu:CDR, emnlp21_CQE, Mao:COTED, Mo:CQRKDD}.
% Therefore, we can assume that the session encoder and the ad-hoc query encoder share a similar embedding space, which is ideal for us to achieve more interpretable session embedding inversion as well as maintain its original retrieval effectiveness. \\

\noindent \textbf{Interpretable query generation. } %Application of Vec2Text
For a session encoder $E_\text{s}$ fine-tuned from an ad-hoc query encoder $E_\text{q}$, we train a Vec2Text model $\phi_{\text{q}}$ based on $E_\text{q}$ but not based on $E_\text{s}$.
Then, for a session embedding $\mathbf{s_i} = E_\text{s}(q_i, H_i)$, we obtain its transformed text $\hat{q_i}=\phi_{\text{q}} (\mathbf{s}_i)$ through $\phi_{\text{q}}$.
Specifically, Vec2Text includes two models: the inversion model and the correction model, and the generation process of Vec2Text includes two steps:
(1) The initial inversion step, where an inversion model first inverts the embedding into an initial inverted text $t^{\text{inv}}$. (2) The correction step, where a correction model then progressively refines this initial inverted text $t^{\text{inv}}$ to be more accurate.
Figure~\ref{model1} shows an illustration of the whole generation process of Vec2Text. The detailed introduction of our Vec2Text model training is provided in Appendix~\ref{vec2text_training}.

Since $E_\text{s}$ and $E_\text{q}$ share the same retrieval embedding space, the transformed query embedding $E_{\text{q}}(\hat{q_i})$ is supposed to be highly similar to the original session embedding $\mathbf{s_i}$ and thus keep similar retrieval performance.

% First, it is worth noting that it is common to initialize the conversational session encoder and the passage encoder from a pre-trained ad-hoc retriever and only fine-tune the session encoder while freezing the passage encoder for facilitating the training~\cite{Yu:CDR, emnlp21_CQE, Mao:COTED, Mo:CQRKDD}.
% We skillfully leverage this training characteristic to achieve interpretable session embedding inversion as well as maintain the original retrieval effectiveness.

% Therefore, to maintain the retrieval effectiveness, we resort to maximizing the similarity between the transformed query embedding $E_a(\hat{q})$ and the session embedding $\mathbf{s}$ as an approximation:
% \begin{eqnarray}
%     \hat{q} \leftarrow \text{transform}(\mathbf{s}), \\
%     \text{maximize}: \quad \text{cos}(E_a(\hat{q}), \mathbf{s}),
% \end{eqnarray}
% where $E_a$ is the ad-hoc query encoder originally paired with the passage encoder $E_p$.

\subsubsection{Interpretability Enhancement with Conversational Query Rewriting}
While the transformed text $\hat{q_i}$ can attain retrieval performance comparable to that of the original session embedding $\mathbf{s_i}$ when encoded by the ad-hoc query encoder $E_q$, there is no assurance that $\hat{q_i}$ will form a coherent and interpretable sentence for human understanding.

We propose a simple method to leverage external query rewrites to enhance the interpretability.
Specifically, we first employ a conversational query rewriting model $R$ (for example, the T5QR~\citep{Sheng-Chieh:CQR} model) to transform the conversational search session $\{q_i, H_i\}$ into a standalone query rewrite $q^{*}_i=R(q_i, H_i)$.
Then, in the generation process of Vec2Text, we discard the initial inversion process and directly use the query rewrite $q^{*}_i$ as the initial inverted text $t^{\text{inv}}$.

The rewriting model $R$, trained on a vast dataset of human-crafted rewrites, ensures that the resultant query rewrite is coherent and understandable compared to the original inverted text produced by VecText's inversion model.
% Since the rewriting model $R$ is learned from a large number of human-written rewrites, the generated query rewrite is  than the originally inverted text generated by the inversion model of VecText.
The new inverted text $q^{*}_i$, serving as an improved starting point for the session embedding transformation, can help lead the whole generation process towards a more interpretable direction, and thus enhance the interpretability of the final transformed text $\hat{q_i}$.

% The new inverted text $q^{*}_i$, serving as a better start point for the session embedding transformation, can effectively improve the quality of the final transformed text $\hat{q_i}$.

% Although the inverted text $\hat{q_i}$ can achieve similar retrieval performance as the original session embedding $\mathbf{s_i}$ by using the ad-hoc query encoder $E_q$, it is not guaranteed that $\hat{q_i}$ is a coherent query sentence that has decent interpretability for human to understand.

\section{Experimental Settings}
This section presents our basic experimental settings. See Appendix~\ref{more_detailed_experimental_settings} for full details.

\subsection{Datasets}
We use four public conversational search datasets: 
QReCC~\citep{qrecc}, TREC CAsT-19~\citep{cast19}, TREC CAsT-20~\citep{cast20}, and TREC CAsT-21~\citep{cast21}.
The QReCC dataset consists of 13.6K conversations, with an average of 6 turns per conversation.
While the three CAsT datasets (19, 20, 21) only comprise 50, 25, and 26 conversations, respectively, but with more detailed relevance labeling.
All four datasets provide human rewrites for each turn.
Following existing works~\citep{LeCoRE, ConvGQR}, we train CDR models on the QReCC dataset and conduct evaluations on the three CAsT datasets.
% We show the detailed dataset introduction in Appendix~\ref{}.
% The basic dataset statistics are shown in Table~\ref{statistics_three_datasets}.

% QReCC~\citep{qrecc}. The conversational dense retriever model is trained on the QReCC training set, with evaluations conducted on the CAsT-19, CAsT-20, CAsT-21, and test datasets from QReCC. Each of the CAsT datasets (2019, 2020, 2021) comprises 50, 25, and 26 conversational search sessions, respectively, with approximately ten conversational queries in each session. The QReCC dataset consists of 13.6K conversations, with an average of 6 turns per conversation. All four datasets provide manually rewritten queries for every question and include relevance judgments for the majority, offering a comprehensive evaluation framework for conversational search systems.

\subsection{Conversational Dense Retrieval Models} 
Currently, there are mainly two paradigms to train conversational session encoders. The first is proposed by \citet{Yu:CDR} which employs an ad hoc query encoder as the teacher and learns the student session encoder by mimicking the teacher embeddings originating from human queries.
The second is to use the classical ranking loss function~\citep{emnlp20_dpr, emnlp21_CQE} to maximize the distance between the session and its positive passages and minimize the distance between the session and negative passages.

Our evaluation is based on both types of CDR models. We name the first type KD-\textit{Retriever} and the second type Conv-\textit{Retriever}, where \textit{Retriever} can be replaced with any base ad-hoc retriever.
Specifically, we mainly experiment with a popular ad-hoc retriever, i.e., GTR~\citep{GTR}, and we investigate the universality of our method to different ad-hoc retrievers in Section~\ref{Investigation_of_different_ad_hoc_retrievers}.

% The selection of Conversational Dense Retriever models involves three ad-hoc retrievers as base models: ANCE~\citep{ANCE}, GTR~\citep{GTR}, and BGE~\citep{BGE}, along with two training methods mentioned in Section~\ref{Our Method} : contrastive learning and knowledge distillation. By leveraging these combinations, we derived six distinct CDR models, namely conv-gtr, KD-GTR, conv-ance, gtr-ance, conv-bge, and kd-bge. For instance, conv-gtr model signifies the use of the GTR model as the base model, with training conducted through contrastive learning.

\subsection{Baselines}

\begin{figure}[!t]
  \centering\includegraphics[width=0.5\textwidth]{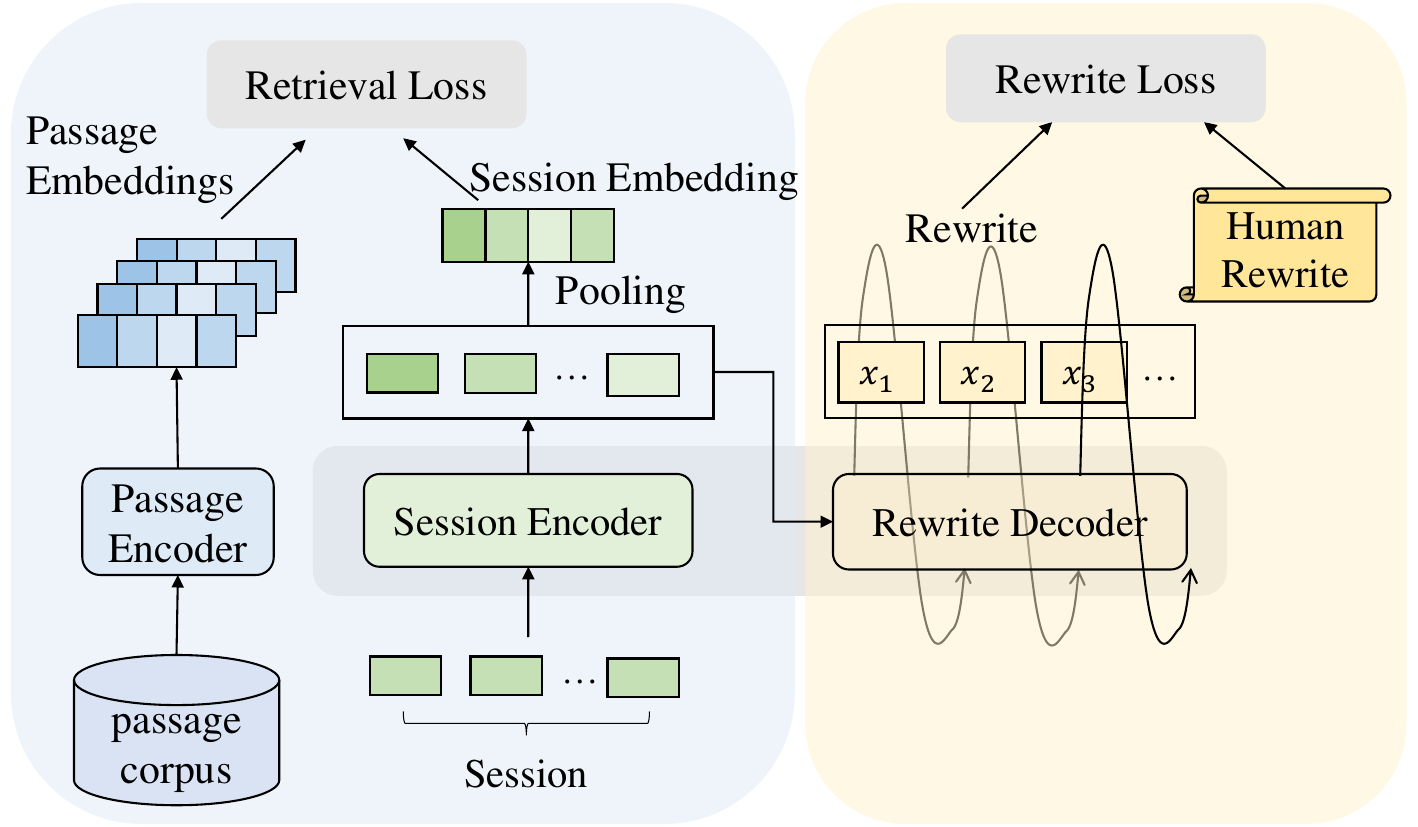}
  \caption{The workflow of UniCRR (\textbf{Uni}fying \textbf{C}onversational Dense \textbf{R}etrieval and Query \textbf{R}ewriting).}
  \label{UniCRR}
\end{figure}

Our main goal is to demonstrate the interpretability and preserved retrieval performance of the transformed text generated by our \ours, compared to the original session embeddings of \textbf{KD-GTR} and \textbf{Conv-GTR}.
To the best of our knowledge, there is no existing method that is completely suitable for our task, i.e., interpreting conversational session embeddings (see the full task definition in Section~\ref{sec:task_forumation}).
Therefore, we propose a straightforward but strong baseline called \textbf{UniCRR}.
Figure~\ref{UniCRR} illustrates UniCRR. Specifically, we unify the session encoder and the query rewriter in an encoder-decoder architecture and adopt multi-task learning to simultaneously train both. 
As such, the rewrite generated from the decoder part can interpret the session embedding generated from the encoder part to some extent. 
% More training details of UniCRR are provided in Appendix~\ref{}.

% We compare our method with four baseline: 

% (1) \textbf{encoder-decoder baseline}: Given the disparities in experimental settings and model architecture between our study and existing works, a direct comparison would be unfair. Consequently, we propose an \textbf{encoder-decoder baseline} to ensure a comparison under similar conditions. We choose to use multi-task learning to simultaneously train both the encoder and decoder components. In the encoder training phase, similar to our method for training the conversational dense retriever, we utilize methods such as knowledge distillation or contrastive learning, and the loss in this training process is termed as $L_{enc}$. Simultaneously, in the decoder training phase, we use human rewrites as labels  for supervised learning, and the loss in this training process is termed as $L_{dec}$. Therefore, the overall model loss is calculated as:
% \begin{equation}
% L =  \sigma \cdot L_{enc} + L_{dec}
% \end{equation}
% where $\sigma$ represents loss weight. Through this approach, we can optimize both the encoder and decoder concurrently, aiming to achieve enhanced performance.

In addition to the original {KD-GTR}, {Conv-GTR}, and our proposed UniCRR, we also use the following conversational search baselines mainly for the comparisons of retrieval performance:
(1) \textbf{T5QR}~\citep{Sheng-Chieh:CQR}: A conversational query rewriter based on T5~\citep{T5}, trained using human-generated rewrites.
(2) \textbf{ConvGQR}~\citep{ConvGQR}: A framework for query reformulation that integrates query rewriting with generative query expansion.
(3) \textbf{LeCoRE}~\citep{LeCoRE}: A conversational lexical retrieval model extending from the SPLADE model with two well-matched multi-level denoising approaches.

\subsection{Evaluation Metrics}
 \textbf{Retrieval and inversion evaluation.}
Following existing works~\citep{ ConvGQR,LeCoRE} and the official settings of the CAsT datasets~\citep{cast20}, we choose MRR, NDCG@3, and Recall@100 to evaluate the retrieval performance. 
We use two metrics to quantify the fidelity of the embedding inversion: (1) The absolute difference in the retrieval performances between using the session embeddings and the transformed text. (2) Following Vec2Text~\citep{vec2text}, we also calculate the cosine similarity between the session embeddings and the transformed text embeddings.

\noindent \textbf{Interpretability evaluation.}
We conduct a human evaluation for the interpretability of the transformed text from three aspects:
(1) \textit{Clarity}: evaluating the clarity of text expression and identifying the presence of ambiguity or vague expressions; (2) \textit{Coherence}: examining the logical structure of the text; (3) \textit{Completeness}: determining the extent to which the text comprehensively covers all historical information. 
% Examples of these three metrics are provided in Appendix~\ref{human_evluation_metrics}.
Five information retrieval researchers are employed to assign scores ranging from 1 to 5. A larger score indicates better performance.

\begin{table*}[!t]
\centering
\setlength\tabcolsep{2pt}
\scalebox{0.83}{\begin{tabular}{c|ccc|ccc|ccc}
\toprule
& \multicolumn{3}{c|}{CAsT-19}& \multicolumn{3}{c|}{CAsT-20}   & \multicolumn{3}{c}{CAsT-21}  \\ \cline{2-10} 
\multirow{-2}{*}{Method} & MRR & NDCG@3  & Recall@100  & MRR    & NDCG@3 & Recall@100   & MRR & NDCG@3 & Recall@100    \\ \midrule
T5QR & 65.8& 41.9    & 38.2   & 46.6   & 32.1   & 41.4    & 47.9& 34.1   & 45.2\\
ConvGQR   & 66.7& 39.3    & 33.7   & 39.7   & 25.9   & 33.8    & 40.6& 25.3   & 37.3\\
LeCoRE    & 70.3& 42.2    & \textbf{49.4}    & 45.0   & 29.0   & 46.7    & 54.8 & 32.3   & 38.7\\ \hline
Conv-GTR  & 53.8& 31.0    & 34.6   & 27.9   & 18.4   & 31.8    & 42.2& 28.4   & 46.4\\
UniCRR    &\gc54.4 (+0.6)&\gc31.9  (+0.9)&\rc31.0  (-3.6)&\rc36.0  (+8.1)&\rc23.7  (+5.3) &\rc33.3 (+1.5)& 
\rc35.0 (-7.2)& 
\rc23.2  (-5.2)& 
\rc31.3  (-15.1) \\
\ours & 
\rc56.4 (+2.6)& 
\rc33.1 (+2.1) & 
\gc37.0 (+2.4) & 
\gc27.2 (-0.7) &\gc18.5 (+0.1) &\gc30.4 (-1.4) &\gc41.9 (-0.3)&\gc28.2 (-0.2)&\gc41.7 (-4.7)  \\ \hline
KD-GTR    & \textbf{74.9} & \textbf{46.9}& 41.9   & \textbf{49.5}    & \textbf{35.9}    & \textbf{46.9}& \textbf{54.7}& 36.4   & \textbf{55.4} \\
UniCRR    &\rc65.1 (-9.8)&\rc40.6 (-6.3) &\rc37.0 (-4.9)&\rc44.4 (-5.1)&\rc32.3 (-3.6)&\rc39.5 (-7.4) &\rc41.0 (-13.7) &\rc27.3 (-9.1)&\rc39.5 (-15.9) \\
\ours   &\gc74.2 (-0.7)&\gc44.9 (-2.0) &\gc43.0 (+1.1)&\gc47.6 (-1.9)&\gc34.4 (-1.5)&\gc44.0 (-2.9) &\gc\textbf{54.7} (+0.0)&\gc\textbf{37.4} (+1.0) &\gc55.1 (-0.3)  \\ \bottomrule
\end{tabular}}
\caption{Retrieval performance comparisons. Our main competitor is UniCRR. The numbers in parentheses indicate the absolute difference between the original CDR model (i.e., Conv-GTR or KD-GTR) and the transformed text. In the comparison between \ours and UniCRR, a green background indicates that its performance gap with the original session embedding is smaller compared to its counterpart, while a red background indicates a larger gap.
The best performance is bold.}
\label{table:retrieval_comparison}
\end{table*}

\begin{table*}[!t]
\centering
\setlength\tabcolsep{2pt}
\scalebox{0.82}{\begin{tabular}{c|ccc|ccc|ccc}
\toprule
& \multicolumn{3}{c|}{CAsT-19}& \multicolumn{3}{c|}{CAsT-20}   & \multicolumn{3}{c}{CAsT-21}  \\ \cline{2-10} 
\multirow{-2}{*}{Method} & MRR & NDCG@3  & Recall@100  & MRR    & NDCG@3 & Recall@100   & MRR & NDCG@3 & Recall@100    \\ \midrule
Conv-GTR  & 53.8& 31.0    & 34.6   & 27.9   & 18.4   & 31.8    & 42.2& 28.4   & 46.4\\

TX-Inversion & 58.0(+4.2) & 33.5 (+2.5)& 37.1(+2.5) & \gc28.2(+0.3) & 18.8(+0.4) & 29.5(-2.3) & 40.7(-1.5) & 26.6(-1.8) &\gc 43.1(-3.3) \\

TX-Human & \gc55.6(+1.8) & \gc33.0(+2.0) & \gc36.0(+1.4) & 27.3(-0.6) & \gc18.5(+0.1) & \gc30.9(-0.9) &42.6(+0.4) & 26.5(-1.9) & 41.1(-5.3) \\

\ours{}   & 
56.4 (+2.6)& 
33.1 (+2.1) & 
37.0 (+2.4) & 
27.2 (-0.7) &\gc18.5 (+0.1) & 30.4 (-1.4) &\gc41.9 (-0.3)&\gc28.2 (-0.2)& 41.7 (-4.7)  \\ \hline
KD-GTR    & \textbf{74.9} & \textbf{46.9}& 41.9   & \textbf{49.5}    & \textbf{35.9}    & \textbf{46.9}& \textbf{54.7}& 36.4   & 55.4 \\

TX-Inversion & 71.6(-3.3) & 44.2(-2.7) &\gc42.3(+0.4) & 48.1(-1.4) & 33.6(-2.3) & 44.8(-2.1) &53.8(-0.9) & \gc36.1(-0.3) &\gc \textbf{55.6}(+0.2) \\

TX-Human & 73.1(-1.8) & 44.1(-2.8) & \gc42.3 (+0.4) & \gc48.6(-0.9) & \gc35.0(-0.9) & \gc45.9(-1.0) &53.1(-1.6) & 35.8(-0.6) & 54.3(-1.1) \\

\ours   &\gc74.2 (-0.7)&\gc44.9 (-2.0) & \textbf{43.0} (+1.1)& 47.6 (-1.9)& 34.4 (-1.5)& 44.0 (-2.9) &\gc\textbf{54.7} (+0.0)& \textbf{37.4} (+1.0) & 55.1 (-0.3)  \\ \bottomrule
\end{tabular}}
\caption{Ablation results of the effect of rewriting-enhancement. The numbers in parentheses indicate the difference between the original (i.e., Conv-GTR or KD-GTR) and the transformed text. \ours{} uses T5QR for rewriting enhancement by default. In the comparison between TX-Inversion, TX-Human, and \ours{}, a green background indicates that its performance gap with the original session embedding is the smallest. The best performance is bold.}
\label{ablation_study}
\end{table*}

\subsection{Implementations}
For \ours, we train Vec2Text models on the large-scale MSMARCO~\citep{nips16_msmarco} query and passage collections based on different ad-hoc query encoders.
The inversion model is trained for 50 epochs with a batch size of 128 and the correction model is trained for 100 epochs with a batch size of 200 with 1e-3 learning rate. The maximum sequence length is set to 48.
By default, we use the rewrites generated by T5QR to perform rewriting enhancement.

We train the conversational dense retrieval models on the QReCC dataset. The session encoder is initialized from an ad-hoc query encoder and the passage encoder is frozen during training.
The input of the session encoder is the concatenation of all historical turns and the current query following existing works~\citep{LeCoRE, ConvGQR}.
For KD-Retriever, we follow \citet{Yu:CDR} using the Mean Squared Error (MSE) loss function to perform knowledge distillation.
For Conv-Retriever, we use the contrastive ranking loss function with 48 batch size.
The maximum input lengths of the session encoder and the passage encoder are set to 512 and 384, respectively.
We generally train 2 epochs with 5e-5 learning rate for CDR models. 

\section{Experimental Results}

\subsection{Retrieval and Inversion Evaluation}
Note that our work does not aim to achieve absolutely higher retrieval performance, but rather to faithfully restore the retrieval performance of the original session embeddings, so the main competitor of our \ours is only UniCRR. 
The retrieval performance comparisons on three CAsT datasets are shown in Table~\ref{table:retrieval_comparison} and the similarity is shown in Table~\ref{similarity}.  We find:

\begin{table}[ht]
\centering
\fontsize{8pt}{12pt}\selectfont
\begin{tabular}{p{0.3\linewidth}|ccc}
\hline
Method & CAsT-19 & CAsT-20 & CAsT-21 \\ \hline
UniCRR & 94.10 & 92.80 & 87.90 \\
ConvInv & \textbf{95.80} & \textbf{95.20} & \textbf{94.50} \\ \hline

\end{tabular}
\caption{The similarity between the embeddings of texts generated by UniCRR and \textsc{ConvInv}, and the original session embeddings. The best performance is bold.}
\label{similarity}
\end{table}

(1) Compared to UniCRR, \textbf{\ours achieves superior embedding restoration}. For example, for KD-GTR, the average absolute differences for \ours are 0.87 
(MRR), 1.5 (NDCG@3), and 1.43 (Recall@100), and the average absolute differences for UniCRR are 9.53 (MRR), 6.3 (NDCG@3), and 9.4 (Recall@100). This indicates that the transformed texts generated by \ours are closer to the original session embeddings. This aligns with the restoration similarity, which is shown in Table~\ref{similarity}.
The superior reconstruction performance of ConvInv compared to UniCRR may stem from the fact that UniCRR fails to establish a direct correlation between session embeddings during both the training and inference phases.
% 与UniCRR相比，ConvInv还原的更好。算一下每个指标上，三个数据集上平均的绝对差值是多少写出来（表明我们还原的更贴近原来的session embedding). This is align with 还原similarity，which is shown in Figure/Table. 然后写一下为什么ConvInv的还原效果比UniCRR好（因为UniCRR没有建立和session emb之间的直接联系balabala。。）

(2) We surprisingly notice that \textbf{the transformed text generated by \ours can sometimes even yield slightly better retrieval performance}. For example, on the CAsT-21 dataset, we observe 2.7\% NDCG@3 relative gains over the original session embedding, respectively. This discovery could potentially pave the way for enhancing retrieval efficacy and interpretability through the collaborative optimization of CQR and CDR.
% （惊喜的是）或者就说we notice that，我们ConvINv还原出来的text有时候甚至可以产生略微更好的检索效果。for example on ... （具体数值举例）,这一发现可能有助于Cdr performance transfer to CQR 或者说CQR和CDR的联合优化（从而产生更好的检索效果的同时有更好的可解释性） 提供思路。

\textbf{Ablation study for rewriting enhancement.}
We propose using external query rewrites generated by T5QR to improve the interpretability of transformed text, which matches the original session embedding's retrieval performance but may lack coherence and understandability. Building on this proposition, we compare three types of transformed text to investigate the effect of rewriting enhancement:
(1) using T5QR rewrites for the rewriting enhancement, which is the default \textsc{ConvInv}.
(2) \textit{TX-Human}: using human rewrites for the rewriting enhancement.
(3) \textit{TX-Inversion}: not performing rewriting enhancement (i.e., just using the text generated by the inversion model for the correction step). The ablation results of the retrieval performance of transformed text are shown in Table~\ref{ablation_study}. We observe that the utilization of rewriting enhancement brings the retrieval performance closer to the original. \textbf{Using rewriting enhancement generally leads to stronger overall retrieval performance compared to not}.
% 这里展示我们用T5rewrite， human rewrite, 和不加rewriting enhancement的检索性能对比。列个表/图。 注意与原来CDR models的性能更接近更好，不一定非要绝对值更高。只要说证明用了rewrite比不加rewriting综合看检索效果强一些就行了。毕竟加rewritng enhancement的主要目的是优化interpretability

\begin{figure*}[ht]
  \centering  \includegraphics[width=0.95\textwidth]{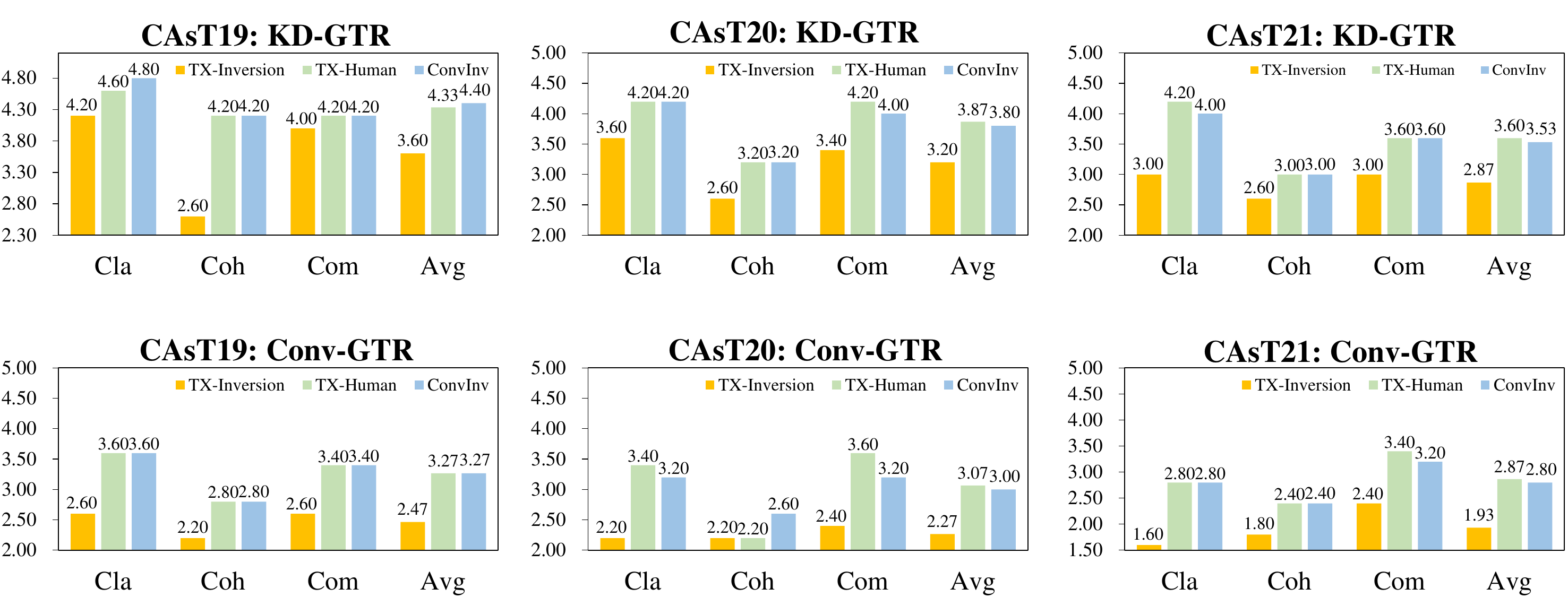}
  \caption{Results of human evaluations for interpretability. Cla, Coh, and Com represent Clarity, Coherence, and Completeness, respectively. The Avg indicates the average of these scores.}
  \label{human_evaluation}
\end{figure*}

\begin{table}[!t]
\centering
\small
\scalebox{0.95}{\begin{tabular}{p{\linewidth}}
\hline\hline
\textbf{Context}: (CAsT-19 Session 54)\\
$q_1$: What is worth seeing in Washington D.C.? \\
... \\
$q_4$: Is the spy museum free?\\
$q_5$: What is there to do in DC after the museums close?\\
\hline
\textbf{Current Query}(68.1):\\
What is the best time to visit the reflecting pools?\\
\hline
\textbf{\ours}(68.1):\\
In Washington D.C. what is the best time to visit the reflecting pools (like the Smithsonian Museum)?\\

\textbf{TX-Human}(47.9):\\
In Washington D.C., what is the best time to visit the reflecting pools by the Smithsonian and other DC museums?\\

\textbf{TX-Inversion}(20.2):\\
In Washington D.C., what is the best time to visit the reflecting pools (e.g. Smithsonian National Museum)?\\
\hline
\textbf{Human Rewrite}(36.1):\\
What is the best time to visit the reflecting pools in Washington D.C.?\\
\hline

\hline\hline
\end{tabular}}
\caption{A case illustrating the distinction in utilizing rewriting enhancement for transformed text. The numbers in parentheses indicate the retrieval performance NDCG@3 of the transformed text. Notably, the number in parentheses under \textbf{Current Query} represents the retrieval results of the original session embedding, not that of the current query statement.}
\label{case_study}
\end{table}

\subsection{Interpretability Evaluation}
We manually evaluate the interpretability of three types of transformed text generated by \textsc{ConvInv}.
Evaluation results are shown in Figure~\ref{human_evaluation} and we have the following observations:

\begin{table*}[!t]
\centering
\setlength\tabcolsep{2pt}
\scalebox{0.76}{\begin{tabular}{c|c|ccc|cc|ccc|cc}
\toprule
\multirow{3}{*}{Retriever}& \multirow{3}{*}{Method} &
\multicolumn{5}{c|}{CAsT-19}& \multicolumn{5}{c}{CAsT-21}  \\ \cline{3-12} 

&& \multicolumn{3}{c|}{Retrieval Performance} & \multicolumn{2}{c|}{Interpretablity} & \multicolumn{3}{c|}{Retrieval Performance} & \multicolumn{2}{c}{Interpretablity}\\ \cline{3-12}

& & MRR & NDCG@3  & Recall@100  & similarity    & hum eval & MRR & NDCG@3  & Recall@100  & similarity    & hum eval   \\ \midrule

\multirow{2}{*}{GTR} & KD-GTR  &\textbf{74.9} & \textbf{46.9}& 41.9 & -  & - & 54.7 & 36.4   & 55.4 & - & -\\

& \ours &74.2 (-0.7)&44.9 (-2.0) & \textbf{43.0} (+1.1) & 0.985 & 4.40 & 54.7(0.0) & 37.4(+1.0) & 55.1(-0.3) & 0.945 & 3.53 \\
\hline

\multirow{2}{*}{ANCE} & KD-ANCE   & 72.0 & 44.4 & 34.2 & - & - & 52.8 & 36.9 & 50.8 & - & -\\

& \ours & 72.0(0.0) & 44.5(+0.1) & 34.3(+0.1) & 0.999 & 4.90 & 55.8(+3.0) & 37.4(+0.5) & 53.1(+2.3) & 0.998 & 4.07 \\
\hline

\multirow{2}{*}{BGE} & KD-BGE   &  69.5 & 44.0 & 41.2 & - & - & 57.9 & \textbf{41.2} & \textbf{56.0} & - & - \\

& \ours & 69.9(+0.4) & 45.4(+1.4) & 41.5(+0.3) & 0.972 & 4.33 & \textbf{59.8}(+1.9) & 41.1(-0.1) & 54.4(-1.6) & 0.954 & 4.25 \\

\bottomrule
\end{tabular}}
\caption{Retrieval performance and interpretability of generated transformed text based on different ad-hoc retrievers on CAsT-19 and CAsT-21 datasets. The "hum eval" represents the human evaluation score. The numbers in parentheses indicate the difference between the original and the transformed text. The best performance is bold.}
\label{investigations_of_different_retrievers}
\end{table*}

% \subsection{Interpretability Evaluation}
% \textbf{Human Evaluation.} Certain key features of interpretability, such as clarity, user-friendliness, and reasonableness, may be challenging to accurately measure using specific numerical or automated evaluation metrics. Human evaluation becomes crucial in capturing these qualitatively nuanced characteristics. Furthermore, automatic search evaluation metrics may not entirely capture the model's capability in comprehending the nuanced contextual intentions behind search queries~\citep{LLM_Know_You}. Last but not least, human evaluators possess the capability to identify error types that automatic evaluation metrics might overlook. These encompass subtle yet significant interpretability issues, such as vague expressions, ambiguity, or contextual errors.

% In assessing the interpretability of inversion text, a scoring system ranging from 1 to 5 is employed, with the following weighted aspects: (1) Clarity: evaluating the clarity of text expression and identifying the presence of ambiguity or vague expressions; (2) Coherence: examining the logical structure of the text; (3) Sensibility: assessing the adherence to common sense and real-world scenarios; (4) Completeness: determining the extent to which the text comprehensively covers all historical information. Each aspect is assigned equal weight. The five evaluators are required to assign scores within the 1 to 5 range based on the specified criteria.

(1) Using query rewrites as the initial inverted text improves the interpretability of the transformed text of KD-GTR and Conv-GTR across the CAsT-19, CAsT-20, and CAsT-21 datasets. This improvement can be attributed to the introduction of the rewrite as the initial inverted text, which essentially offers the corrector model a more informative and clear starting point.
These notable enhancements underscore the necessity of our rewriting-enhancement approach in improving text interpretability.

(2) For both KD-GTR and Conv-GTR, the human evaluation scores of transformed text on CAsT-19 are higher, whether using rewriting-enhancement or not, compared to CAsT-20 and CAsT-21. This observation may be attributed to the absence of response information in the CAsT-19 dataset, which exclusively contains query content. Consequently, the session embedding on CAsT-19 is relatively simple, lacking the complexity introduced by response data. 

%However, on CAsT-20 and CAsT-21, a shift occurs due to the introduction of response information and more complex session embeddings. This heightened complexity introduces ambiguity and polysemy to the text, making it more challenging for transformed text to capture and reconstruct semantic relationships within the context. 

(3) The lower human evaluation scores of transformed text for Conv-GTR compared to KD-GTR on three datasets may be due to the implications of contrastive learning. This method often introduces additional noise. Therefore, Conv-GTR's session embedding might be more prone to interference, potentially leading to its less effective performance in generating transformed text.

We provide a concrete example of the transformed texts in Table~\ref{case_study}. 
More case studies are in Appendix~\ref{additional_case_study}.
We find that 
the transformed text \ours{} not only exhibits high interpretability, fully capturing the user's query intent about ``in Washington D.C.'', but also maintains the closest proximity of retrieval performance to the original session embedding.
We notice that it includes an additional clue ``(like the Smithsonian Museum)'' in the query, which may just be additional knowledge reflected in the mysterious session embedding that can help retrieve passages about famous attractions in Washington D.C.

\subsection{Experiments with Different Retrievers}
\label{Investigation_of_different_ad_hoc_retrievers}
We investigate the universality of our \ours by changing the base ad-hoc retriever of the CDR models.
Specifically, we experiment with another two popular ad-hoc retrievers: ANCE~\citep{ANCE} and BGE~\citep{BGE}. 
Results are shown in Table~\ref{investigations_of_different_retrievers}.
% The retrieval and interpretability evaluations on CAsT-19 and CAsT-21 datasets are shown in Table~\ref{investigations_of_different_retrievers}, and more results are provided in Appendix~\ref{investigation_different_retrievers_cast19_cast20_cast21}. 
We find that:

(1) Regardless of the selected ad-hoc retriever, both retrieval similarity and text similarity metrics are observed to be high. To illustrate, on the CAsT-19 dataset, the average absolute differences for KD-ANCE on CAsT-19 dataset are 0.0 
(MRR), 0.1 (NDCG@3), and 0.1 (Recall@100), and the cosine similarity is up to 99.9\%.

(2) Across both CAsT-19 and CAsT-21 datasets, there is a sustained consistency between similarity scores and human evaluations, indicating that textual similarity is a reliable indicator of quality as perceived by human judges. However, this does not encapsulate all the factors considered in human evaluations, especially as similarity scores remain robust while human evaluations show a decline from CAsT-19 to CAsT-21. 
% The \textsc{ConvInv} method with the GTR retriever, for example, achieves a similarity of 0.945 and is accompanied by a corresponding robust human evaluation score of 3.53. 
Although there is a noted decrease in human evaluation scores across all methods when moving from CAsT-19 to CAsT-21, the similarity scores remain high or even show marginal improvement. 

\section{Conclusion}

In this paper, we present a novel approach \textsc{ConvInv} to shed light on the interpretability of conversational dense retrieval.
% Our approach involves transforming session embeddings into explicit and interpretable text, providing an intuitive understanding of the distinctive characteristics of various CDR models. 
By experimenting with two typical conversational dense retrieval models on three conversational search benchmarks, we demonstrate the effectiveness of our approach in providing interpretable text as well as faithfully restoring the original retrieval performance of session embeddings.
Our work not only enhances interpretability in conversational dense retrieval but also lays a groundwork for future research toward trustworthy conversational search.

\section*{Limitations}
Our work provides a simple but effective solution to enhance the interpretability of conversational dense retrieval models, bridging the gap between opaque session embeddings and transparent query rewriting. However, the necessity to train distinct Vec2Text models based on various retrievers demands a significant time investment. Additionally, for session embeddings trained using contrastive learning, the transformed text fails to achieve sufficiently high similarity to the original session embedding, suggesting an incomplete decoding of the session embedding. Besides, some of the transformed texts may not exhibit retrieval performance as effective as the original session embeddings.
Some more sophisticated conversational dense retrievers have not been investigated.

\section*{Acknowledgments}
This work was supported by National Key R\&D Program of China No. 2022ZD0120103, National Natural Science Foundation of China No. 62272467,  the fund for building world-class universities (disciplines) of Renmin University of China, and Public Computing Cloud, Renmin University of China. The work was partially done at the Engineering Research Center of Next-Generation Intelligent Search and Recommendation, MOE.

\bibliography{custom}

\appendix

\section{Appendix}
\label{sec:appendix}
\begin{table}[!t]
\centering
\scalebox{0.8}{
\begin{tabular}{ccc@{ }c@{ }c@{ }c}
\hline
\multirow{2}{*}{Statistics} & \multicolumn{2}{c}{QReCC}& CAsT-19 & CAsT-20 & CAsT-21 \\ 
 \cmidrule(r){2-3}  \cmidrule(r){4-4}  \cmidrule(r){5-5}  \cmidrule(r){6-6}

& Train & Test & Test & Test & Test \\ \hline
\# Conv. & 10823 & 2775 & 50 & 25 & 26 \\ 
\# Questions & 63501 & 16451 & 479 & 216 & 239 \\ 
\# Documents & \multicolumn{2}{c}{54M} & 38M & 38M & 40M \\ \hline
\end{tabular}}
\caption{Data statistics of conversational search datasets.}
\label{statistics_four_datasets}
\end{table}

\subsection{Vec2Text}
\label{vec2text_training}
Due to the necessity of transforming session embeddings into explicit and interpretable text, we integrate the Vec2Text model into our architecture. The utilization of Vec2Text~\citep{vec2text} is driven by its capability to effectively invert the full text represented in dense text embeddings, aligning with our goal to provide interpretability of session embeddings in conversational dense retrieval.

The Vec2Text model aims for the complete inversion of input text from its embedding; it leverages the difference between a hypothesis embedding and a ground-truth embedding to make discrete adjustments to the text hypothesis. Specifically, the Vec2Text model begins by proposing an initial hypothesis and subsequently refines this hypothesis through iterative corrections. The goal is to progressively bring the hypothesis's embedding $\hat{e}^{t} $ closer to the target embedding $e$.

The Vec2Text comprises two models: the inversion model and the corrector model. Firstly, the inversion model endeavors to invert encoder $\phi$ by learning a distribution of texts given embeddings $p\left ( x\mid e,\theta  \right ) $. The training objective for the inversion model is to find $\theta $ using maximum likelihood estimation:
\[ \theta = arg\max_{\hat{\theta}  }E_{x\sim D} \left [ p\left ( x\mid \phi \left ( x \right );\theta   \right )  \right ] \]
On the basis of the simple learned inversion hypothesis $x^{0}$, the corrector model iteratively refines this hypothesis via marginalizing over intermediate hypotheses:
{\small
\[p\left ( x^{\left ( t+1 \right ) }  \mid e\right )   = \sum_{x^{\left ( t \right ) }}p\left ( x^{\left ( t \right ) }  \mid e\right )p\left ( x^{\left ( t+1 \right ) }  \mid e,x^{\left ( t \right ) },\hat{e}^{\left ( t \right ) } \right ) \]
}
where $\hat{e}^{\left ( t \right ) }=\phi \left ( x^{\left ( t \right ) }  \right ) $.

\begin{table}[t]
\centering
\scalebox{1.0}{\begin{tabular}{p{\linewidth}}
\hline\hline
\textbf{Context}: (CAsT-19 Session 79)\\
$q_1$: What is taught in sociology? \\
$q_2$: What is the main contribution of Auguste Comte? \\
$q_3$: What is the role of positivism in it? \\ 
$q_4$: What is Herbert Spencer known for?\\
$q_5$: How is his work related to Comte?\\
\hline
\textbf{Current Query}(35.2):\\
What is the functionalist theory?\\
\hline
\textbf{\ours}(46.9):\\
what is comte's functionalist theory in philosophy?\\

\textbf{TX-Human}(46.9):\\
what is comte's functionalist theory in philosophy?\\

\textbf{TX-Inversion}(20.7):\\
What is the functionalist theory?\\
\hline
\textbf{Human Rewrite}(38.3):\\
What is the functionalist theory in sociology?\\
\hline

\hline\hline
\end{tabular}}
\caption{An additional case illustrating the distinction in utilizing rewriting enhancement for transformed text. The numbers in parentheses indicate the retrieval performance NDCG@3 of the transformed text. Notably, the number in parentheses under \textbf{Current Query} represents the retrieval results of the original session embedding, not that of the current query statement.}
\label{more_case_study}
\end{table}

\subsection{More Detailed Experimental Settings}
\label{more_detailed_experimental_settings}

\subsubsection{Details of Datasets}
The statistical data for each dataset are presented in Table~\ref{statistics_four_datasets} and a more detailed description is provided as follows:

\textbf{QReCC} is a large dataset designed for the study of conversational search. Every query is accompanied by an answer and a human-generated rewrite. QReCC includes a total of 13,598 dialogues featuring 79,952 queries. Of these, 9.3K conversations originate from QuAC questions; 80 from TREC CAsT; and 4.4K from NQ. Additionally, 9\% of the questions within QReCC lack corresponding answers.

\textbf{CAsT-19}, \textbf{CAsT-20}, and \textbf{CAsT-21} are three widely used conversational search datasets released by TREC Conversational Assistance Track (CAsT). For CAsT-19, relevance assessments are available for 173 queries within 20 test conversations. For CAsT-20, the majority of queries are accompanied by relevance judgments. For CAsT-21, there are relevance judgments for 157 queries within 18 test conversations. CAsT-19 and CAsT-20 share the same corpus, whereas CAsT-21 employs a different one.

\subsubsection{Implementation Details}
\label{implementation_details}
During the training process, we conduct the training experiments of the Vec2Text model on four Nvidia A100 40G GPUs. We use bf16 precision and AdamW optimizer with 0.001 as the initial learning rate. The strategy to adjust the learning rate is constant with warm-up. We choose T5~\citep{T5} as the backbone model. The number of times to repeat embedding along the T5 input sequence length is set to 16.

During the inference process, the sequence beam width and the invert num steps are set to 10 and 30, respectively. The maximum input length and the maximum response length are set to 512 and 100, respectively. The dense retrieval is performed using Faiss~\citep{Faiss}.

\subsection{Examples of Human Evaluation}
Examples of the three metrics for human evaluation are shown in Table~\ref{human_evluation_metrics}.

\subsection{Supplement of Case Study}
\label{additional_case_study}
In this section, We provide an additional case in Table~\ref{more_case_study} for analysis. The transformed text not only includes the keyword of the original query "functionalist theory", but also enriches it with additional information "comte" and "philosophy", thus yielding a retrieval performance that surpasses that of the human rewrite.

\subsection{Experiments with Different Retrievers}
\label{investigation_different_retrievers_cast19_cast20_cast21}

Investigations of Based on Different Ad-hoc Retrievers on CAsT-19, CAsT-20, and CAsT-21 datasets are shown in Table~\ref{investigation_different_retrievers_cast19}, Table~\ref{investigation_different_retrievers_cast20} and Table~\ref{investigation_different_retrievers_cast21}, separately.

\begin{table*}[ht]
\begin{center}
\begin{tabular}{cl}
\hline
\multirow{11}{*}{\textbf{Clarity}} & \textit{\textbf{Context:}} \\  
 & $q_1$: What is throat cancer? \\  
 & $q_2$: Is it treatable? \\ 
 & $q_3$: Tell me about lung cancer. \\ 
 & $q_4$: What are its symptoms? \\ 
 & $q_5$: Can it spread to the throat? \\ 
 & $q_6$: What causes throat cancer? \\ 
 \cline{2-2} 
 & \textit{\textbf{Query:}} What is the first sign of it? \\ 
 & \textit{\textbf{Human Rewrite:}} What is the first sign of throat cancer? \\ 
 & \textit{\textbf{Positive Example:}}What is throat cancer and what is the first sign of it? \\  
 & \textit{\textbf{Negative Example:}} what is the first sign of throat or lung cancer? \\ \hline

 \multirow{8}{*}{\textbf{Coherence}} & \textit{\textbf{Context:}} \\  
 & $q_1$: What are the different types of sharks? \\  
 & $q_2$: Are sharks endangered?  If so, which species? \\ 
 & $q_3$: Tell me more about tiger sharks. \\ 
 \cline{2-2} 
 & \textit{\textbf{Query:}} What is the largest ever to have lived on Earth? \\ 
 & \textit{\textbf{Human Rewrite:}} What is the largest shark ever to have lived on Earth? \\ 
 & \textit{\textbf{Positive Example:}}What's the largest sharks to have ever lived on earth? \\  
 & \textit{\textbf{Negative Example:}} What is the largest ever to have lived on earth, shark sharks? \\ \hline

  \multirow{8}{*}{\textbf{Completeness}} & \textit{\textbf{Context:}} \\  
 & $q_1$: What are the origins of popular music? \\  
 & $q_2$: What are its characteristics? \\ 
 & $q_3$: What technological developments enabled it? \\ 
 \cline{2-2} 
 & \textit{\textbf{Query:}} When and why did people start taking pop seriously? \\ 
 & \textit{\textbf{Human Rewrite:}} When and why did people start taking pop music seriously? \\ 
 & \textit{\textbf{Positive Example:}}When did people start taking pop music seriously. and why? \\  
 & \textit{\textbf{Negative Example:}} What causes pop music and when did it begin to be taken seriously? \\ \hline
 
\end{tabular}
\caption{Examples of the criteria of three metrics of human evaluation.}
\label{human_evluation_metrics}
\end{center}
\end{table*}

\begin{table*}[ht]
\centering
\scalebox{0.83}{\begin{tabular}{c|c|ccc|cc}
\hline
\multirow{2}{*}{Method}  & \multirow{2}{*}{Retriever} & \multicolumn{3}{c|}{Retrieval Performance} & \multicolumn{2}{c}{Interpretability}  \\
\cline{3-7}  
 & & MRR & NDCG@3 &  Recall@100 & similarity & human evaluation \\

\hline
\multirow{6}{*}{KD} & KD-GTR  & \textbf{74.9} & \textbf{46.9} & 41.9 & - & - \\ 
& \ours & 74.2(-0.7) & 44.9(-2.0) & \textbf{43.0}(+1.1) & 0.958 & 4.40 \\
\cline{2-7} 
& KD-ANCE &  72.0 & 44.4 & 34.2 & - & -\\ 
& \ours & 72.0(0.0) & 44.5(+0.1) & 34.3(+0.1) & 0.999 & 4.90 \\ 
\cline{2-7} 
& KD-BGE &  69.5 & 44.0 & 41.2 & - & - \\ 
& \ours & 69.9(+0.4) & 45.4(+1.4) & 41.5(+0.3) & 0.972 & 4.33 \\
\hline
\multirow{6}{*}{Conv} & Conv-GTR &  53.8 & 31.1 & 34.6 & - & - \\
& \ours & 56.4(+2.6) & 33.1(+2.0) & 37.0(+2.4) & 0.778 & 3.27 \\
\cline{2-7}
& Conv-ANCE & 62.8 & 34.5 & 29.6 & - & -\\
& \ours & 47.6(-15.2) & 27.2(-7.3) & 22.0(-7.6) & 0.974 & 4.13 \\
\cline{2-7}
& Conv-BGE & 59.6 & 35.1 & 36.4 & - & -\\
& \ours & 55.2(-4.4) & 32.0(-3.1) & 37.1(+0.7) & 0.736 & 3.47 \\
\hline
\end{tabular}}
\caption{Retrieval performance and interpretability of generated transformed text based on different ad-hoc retrievers on CAsT-19 Dataset. The best performance is bold.}
\label{investigation_different_retrievers_cast19}
\end{table*}

\begin{table*}
\centering
\scalebox{0.83}{\begin{tabular}{c|c|ccc|cc}
\hline
\multirow{2}{*}{Method}  & \multirow{2}{*}{Retriever} & \multicolumn{3}{c|}{Retrieval Performance} & \multicolumn{2}{c}{Interpretability}  \\
\cline{3-7}  
 & & MRR & NDCG@3 & R@100 & similarity & human evaluation \\

\hline
\multirow{6}{*}{KD} & KD-GTR & 49.5 & \textbf{35.9} & \textbf{46.9} & - & - \\
& \ours & 47.6(-1.9) & 34.4(-1.5) & 44.0(-2.9) & 0.952 & 3.80 \\

\cline{2-7} 
&  KD-ANCE  & \textbf{51.0} & 35.8 & 38.6 & - & -\\
& \ours & 49.2(-1.8) & 34.1(-1.7) & 39.9(+1.3) & 0.999 & 4.60 \\

\cline{2-7} 
& KD-BGE &  44.7 & 31.9 & 46.8 & - & - \\
& \ours & 43.3(-1.4) & 30.5(-1.4) & 45.3(-1.5) & 0.966 & 4.25 \\

\cline{2-7} 
\hline
\multirow{6}{*}{Conv} & Conv-GTR &  27.9 & 18.4 & 31.8 & - & - \\
& \ours & 27.2(-0.7) & 18.5(+0.1) & 30.4(-1.4) & 0.719 & 3.00 \\

\cline{2-7}
&  Conv-ANCE & 38.4 & 25.8 & 31.5 & - & -\\
& \ours & 27.8(-10.6) & 18.6(-7.2) & 22.8(-8.7) & 0.972 & 2.93 \\

\cline{2-7}
&  Conv-BGE &  30.7 & 20.9 & 35.4 & - & -\\
& \ours & 31.5(+0.8) & 21.4(+0.5) & 34.0(-1.4) & 0.733 & 3.13 \\

\hline
\end{tabular}}
\caption{Retrieval performance and interpretability of generated transformed text based on different ad-hoc retrievers on CAsT-20 Dataset. The best performance is bold.}
\label{investigation_different_retrievers_cast20}
\end{table*}

\begin{table*}
\centering
\scalebox{0.83}{\begin{tabular}{c|c|ccc|cc}
\hline
\multirow{2}{*}{Method}  & \multirow{2}{*}{Retriever} & \multicolumn{3}{c|}{Retrieval Performance} & \multicolumn{2}{c}{Interpretability}  \\
\cline{3-7}  
 & & MRR & NDCG@3 & R@100 & similarity & human evaluation \\

\hline
\multirow{6}{*}{KD} & KD-GTR & 54.7 & 36.4 & 55.4 & - & - \\
& \ours & 54.7(0.0) & 37.4(+1.0) & 55.1(-0.3) & 0.945 & 3.53 \\

\cline{2-7} 
&  KD-ANCE  & 52.8 & 36.9 & 50.8 & - & -\\
& \ours & 55.8(+3.0) & 37.4(+0.5) & 53.1(+2.3) & 0.998 & 4.07 \\

\cline{2-7} 
& KD-BGE &  57.9 & \textbf{41.2} & \textbf{56.0} & - & - \\
& \ours & \textbf{59.8}(+1.9) & 41.1(-0.1) & 54.4(-1.6) & 0.954 & 4.25 \\

\cline{2-7} 
\hline
\multirow{6}{*}{Conv} & Conv-GTR &  42.2 & 28.4 & 46.4 & - & - \\
& \ours & 41.9(-0.3) & 28.2(-0.2) & 41.7(-4.7) & 0.664 & 2.80 \\

\cline{2-7}
&  Conv-ANCE  & 41.1 & 25.2 & 42.1 & - & -\\
& \ours & 30.1(-11) & 16.9(-8.3) & 31.2(-10.9) & 0.973 & 2.73 \\

\cline{2-7} 
& Conv-BGE &  48.4 & 32.8 & 51.1 & - & - \\
& \ours & 50.5(+2.1) & 32.4(-0.4) & 50.5(-0.6) & 0.740 & 3.07 \\

\hline
\end{tabular}}
\caption{Retrieval performance and interpretability of generated transformed text based on different ad-hoc retrievers on CAsT-21 Dataset. The best performance is bold.}
\label{investigation_different_retrievers_cast21}
\end{table*}

\end{document}